\documentclass[10pt,twoside,twocolumn,english,aps,manuscript,superscriptaddress,prbt,aps, preprint,manuscript,showpacs]{revtex4}
\usepackage{lmodern}
\usepackage[T1]{fontenc}
\usepackage[latin9]{inputenc}
\usepackage{color}
\usepackage{babel}
\usepackage{amstext}
\usepackage{amssymb}
\usepackage{graphicx}
\usepackage[unicode=true,pdfusetitle,
 bookmarks=true,bookmarksnumbered=false,bookmarksopen=false,
 breaklinks=false,pdfborder={0 0 1},backref=false,colorlinks=false]
 {hyperref}

\makeatletter

\newcommand*\LyXThinSpace{\,\hspace{0pt}}
\newcommand{\lyxmathsym}[1]{\ifmmode\begingroup\def\b@ld{bold}
  \text{\ifx\math@version\b@ld\bfseries\fi#1}\endgroup\else#1\fi}

\providecommand{\tabularnewline}{\\}

\newcommand{\lyxdeleted}[3]{}

\@ifundefined{textcolor}{}
{%
 \definecolor{BLACK}{gray}{0}
 \definecolor{WHITE}{gray}{1}
 \definecolor{RED}{rgb}{1,0,0}
 \definecolor{GREEN}{rgb}{0,1,0}
 \definecolor{BLUE}{rgb}{0,0,1}
 \definecolor{CYAN}{cmyk}{1,0,0,0}
 \definecolor{MAGENTA}{cmyk}{0,1,0,0}
 \definecolor{YELLOW}{cmyk}{0,0,1,0}
}

\@ifundefined{date}{}{\date{}}
\makeatother

\begin{document}

\title{Bose-Einstein condensation of triplons in the $S=1$ tetramer antiferromagnet
K$_{2}$Ni$_{2}$(MoO$_{4}$)$_{3}$: A compound close to quantum
critical point}

\author{B. Koteswararao}

\email{koti.iitb@gmail.com}

\affiliation{School of Physics, University of Hyderabad, Central University PO,
Hyderabad 500046, India}

\affiliation{Center of Condensed Matter Sciences, National Taiwan University,
Taipei 10617, Taiwan}

\author{P. Khuntia}

\affiliation{Max-Planck Institute for Chemical Physics of Solids, 01187 Dresden,
Germany}

\affiliation{Department of Physics, Indian Institute of Technology Madras, Chennai-600036,
India}

\author{R. Kumar}

\affiliation{Department of Physics, Indian Institute of Technology Bombay, Powai,
Mumbai 400076, India}

\affiliation{Tata Institute of Fundamental Research, Homi Bhabha Road, Colaba,
Mumbai- 400005, India}

\author{A.V. Mahajan}

\affiliation{Department of Physics, Indian Institute of Technology Bombay, Powai,
Mumbai 400076, India}

\author{Arvind Yogi}

\affiliation{Tata Institute of Fundamental Research, Homi Bhabha Road, Colaba,
Mumbai- 400005, India}

\author{M. Baenitz }

\affiliation{Max-Planck Institute for Chemical Physics of Solids, 01187 Dresden,
Germany}

\author{Y. Skourski}

\affiliation{Dresden High Magnetic Field Laboratory (HLD-EMFL), Helmholtz-Zentrum
Dresden-Rossendorf, 01328 Dresden, Germany}

\author{F. C. Chou}

\email{fcchou@ntu.edu.tw}

\affiliation{Center of Condensed Matter Sciences, National Taiwan University,
Taipei 10617, Taiwan}

\date{{\today}}
\begin{abstract}
The structure of K$_{2}$Ni$_{2}$(MoO$_{4}$)$_{3}$ consists of
$S=1$ tetramers formed by Ni$^{2+}$ ions. The magnetic susceptibility
$\chi(T)$ and specific heat $C_{P}(T)$ data on a single crystal
show a broad maximum due to the low-dimensionality of the system with
short-range spin correlations. A sharp peak is seen in $\chi(T)$
and $C_{P}(T)$ at about 1.13 K, well below the broad maximum. This
is an indication of magnetic long-range order $i.e.,$ the absence
of spin-gap in the ground state. Interestingly, the application of
a small magnetic field ($H>0.1$ T) induces magnetic behavior akin
to Bose-Einstein condensation (BEC) of triplon excitations observed
in some spin-gap materials. Our results demonstrate that the temperature-field
($T-H$) phase boundary follows a power-law $(T-T_{N})\propto H{}^{1/\alpha}$
with the exponent $1/\alpha$ close to $2/3$, as predicted for BEC
scenario. The observation of BEC of triplon excitations in small $H$
infers that K$_{2}$Ni$_{2}$(MoO$_{4}$)$_{3}$ is located in the
proximity of a quantum critical point, which separates the magnetically
ordered and spin-gap regions of the phase diagram. 
\end{abstract}
\maketitle
Spin-gap materials exhibit remarkably exotic magnetic phenomena such
as the realizations of Bose-Einstein condensation (BEC) and appearance
of magnetization plateaus \cite{T. Giamarchi NP2008,Vivien Zapf RMP(2014),H KageyamaPRL1999,KOnizuka JPSJ2000,KKodamaScience2002}.
In general, spin-gap materials have a singlet ($S=0$) ground state
and the triplet excited states are separated from the ground state
by an energy gap, called the spin-gap. With increasing magnetic field
(which leads to a Zeeman splitting of $S=1$ states), at a critical
value of the field $H_{c}$, the lowest sub-state of the triplet ($S_{z}=1$)
crosses the $S=0$ ground state. As a result, a finite concentration
of triplets (triplons) populate. This consequently leads to several
field-induced magnetic long-range-ordering (LRO) phenomena such as
BEC of triplons in the vicinity of $T=0$ K and $H_{c}$ \cite{T. Giamarchi NP2008,Vivien Zapf RMP(2014)}.
In this context, the applied magnetic field ($H$) acts as a chemical
potential in separating the spin-gap region and LRO region of the
quantum phase diagram at $T\rightarrow0$ K \cite{sachdev2014}. Experimentally,
the field-induced BEC of triplon behavior has been intensively studied
for various spin-gap materials with $S=1/2$ dimers TlCuCl$_{3}$
\cite{T. NikuniPRL2000,PMerchant2014}, BaCuSi$_{2}$O$_{6}$ \cite{Sebastian2005,MJaimePRL2004},
Sr$_{3}$Cr$_{2}$O$_{8}$ \cite{Y. Singh2007,A A AczelPRL2009},
Ba$_{3}$Cr$_{2}$O$_{8}$ \cite{A A Aczel2009}. Recently, BEC of
triplet and quintuplet excitations have also been observed above the
critical fields 8.7 T, and 32.42 T, respectively, in the $S=1$ dimer
compound Ba$_{3}$Mn$_{2}$O$_{8}$ \cite{M. UchidaJPSJ2001,E. C. SamulonPRL2009}.
On the other hand, BEC of magnons has been observed in other class
of materials with magnetic-LRO including, yttrium\textendash iron\textendash garnet
films at room temperature via microwave pumping \cite{SODemokritovNature2006},
Cs$_{2}$CuCl$_{4}$ \cite{TRaduPRL2005} and Gd nanocrystalline samples
\cite{SNKaulPRL2011,SNKual Review}. In the case of Cs$_{2}$CuCl$_{4}$,
although the material undergoes a magnetic transition ($T_{N}$) at
0.595 K, the gap in the magnon spectrum closes at about 8.51 T and
the three dimensional (3D) BEC phase boundary relation $T_{N}\propto(H-H_{c})$$^{1/\alpha}$
with an exponent $1/\alpha=2/3$ is observed, similar to other spin-gap
materials \cite{T. NikuniPRL2000,PMerchant2014,Sebastian2005,MJaimePRL2004,Y. Singh2007,A A AczelPRL2009,A A Aczel2009,M. UchidaJPSJ2001,E. C. SamulonPRL2009}.
Interestingly, when a spin-gap system is subjected to significant
three-dimensional interactions, the triplet states are broadened and
thus reduce the size of the spin-gap. In such a case, a small $H_{c}$
is enough to induce BEC of triplon excitations. This class of material
offers an ideal ground to explore quantum critical phenomena in the
proximity of a Quantum Critical Point (QCP) in view of their collective
spin excitations, high homogeneity in boson density, and topological
order \cite{T. Giamarchi NP2008}.

In this Rapid Communication, we study a new kind of antiferromagnetic
material K$_{2}$Ni$_{2}$(MoO$_{4}$)$_{3}$, which exhibits magnetic
LRO, through the comprehensive thermodynamic studies on single crystals.
Interestingly, it exhibits a field-induced BEC of triplon excitations
at low magnetic fields. Being a non-spin-gap material, this quantum
magnet pose to host exotic magnetic excitations and is located close
to a QCP. 

\begin{figure}[t]
\begin{centering}
\includegraphics[clip,scale=0.34]{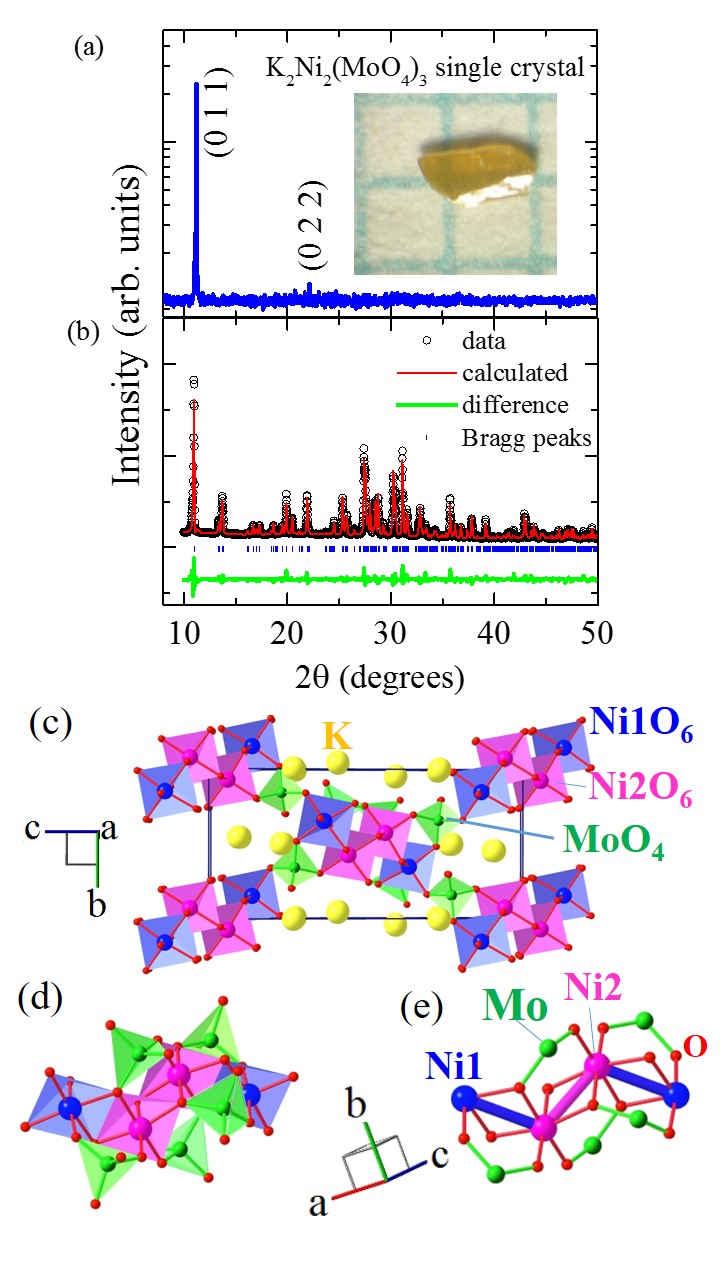}
\par\end{centering}

\caption{\textcolor{black}{(Color online)}\label{1XRD=000026CRYST}\textcolor{black}{{}
(a) XRD pattern of a single crystal of K$_{2}$Ni$_{2}$(MoO$_{4}$)$_{3}$
with the x-ray beam perpendicular to the ($0ll$) planes. Inset shows
an image of the single crystal. The top surface is a ($0ll$) plane.
(b) Rietveld refinement of the XRD data of polycrystalline samples.
(c) The crystal structure of K$_{2}$Ni$_{2}$(MoO$_{4}$)$_{3}$
viewed along $a$-direction. (d) The $S=1$ tetramers are constructed
by Ni1O$_{6}$ (blue) and Ni2O$_{6}$ (pink) octahedra. The MoO$_{4}$
tetrahedral units (green) also mediated the Ni-Ni couplings in a tetramer.
(e) Tetramers in the atoms and bonds representation.}}
\end{figure}

The polycrystalline samples of the titled material were prepared using
K$_{2}$CO$_{3}$, NiO, and MoO$_{3}$. A mixture of these chemicals
with the stoichiometric molar ratio of 1:2:3 was fired for 24 hrs
with a heating rate of 60$^{\circ}$C per hour to reach 600$^{\circ}$C.
The single crystals were grown using K$_{2}$MoO$_{4}$ flux agent
(see inset of Fig. \ref{1XRD=000026CRYST}$(a)$). The x-ray diffraction
(XRD) measurements were done on both the single crystal and polycrystalline
sample. The identified peaks, which correspond to ($0ll$) planes
of the K$_{2}$Ni$_{2}$(MoO$_{4}$)$_{3}$ phase \cite{structureKO12}
are shown in Fig. \ref{1XRD=000026CRYST}$(a)$. In order to extract
the unit cell lattice parameters, we have employed the Rietveld refinement
analysis on the polycrystalline sample with the Fullprof Suite program
\cite{rietveld} using the initial structural parameters provided
by R. F. Klevtsova, $et$ $al.$ in Ref. \cite{structureKO12} (see
Fig. \ref{1XRD=000026CRYST}$(b)$). \textcolor{black}{The obtained
residual refinement factors $R_{P}$, $R_{wp}$, $R_{exp}$, and $\chi^{2}$
are 0.177, 0.180, 0.035, and 5.1, respectively.} The lattice parameters
are found to be $a=6.952(5)$ Å, $b=8.910(7)$ Å, $c=19.740(10)$
Å and $\beta=108.065(5){}^{\circ}$, consistent with the earlier reports
\cite{structureKO12}. 

The compound K$_{2}$Ni$_{2}$(MoO$_{4}$)$_{3}$ crystallizes in
the primitive monoclinic space group $P2_{1}/c$ (No. 14) containing
Z = 4 formula units per unit cell (see Fig. \ref{1XRD=000026CRYST}$(c)$).
The structure has $S=1$ (Ni$^{2+}$) tetramers formed by two edge-shared
Ni1O$_{6}$ and Ni2O$_{6}$ octahedra (see Fig. \ref{1XRD=000026CRYST}$(c)$
and $(d)$). The bond angles of Ni-O-Ni are in between 94$^{\circ}$to
98$^{\circ}$, which naively suggests that the magnetic couplings
might be antiferromagnetic in nature. In a tetramer unit, the Ni$^{2+}$
ions are connected $via$ MoO$_{4}$ tetrahedra, which might lead
to the magnetic frustration through the next nearest neighbor (nnn)
interactions in the tetramer. These $S=1$ tetramers are also connected
to each other through MoO$_{4}$ tetrahedral units running in all
the three crystallographic directions, suggesting the presence of
non-negligible three-dimensional (3D) interactions. 

\begin{figure}
\begin{centering}
\includegraphics[scale=0.36]{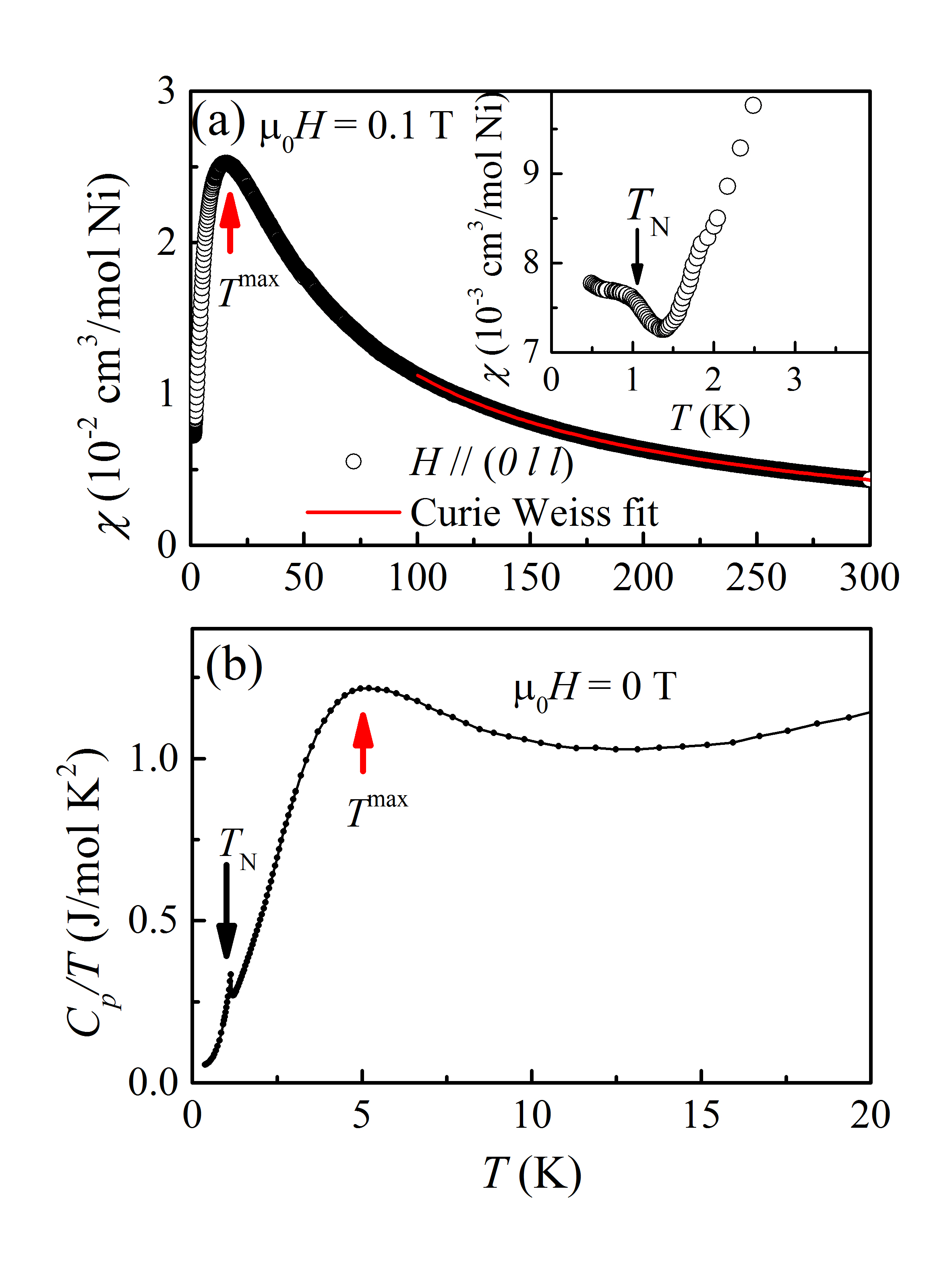}
\par\end{centering}

\caption{\textcolor{black}{(Color online) }\label{2ChiCp}\textcolor{black}{{}
(a) }$\chi$($T$) of a single crystal with $H//(0ll)$. The red line
is a fit to the Curie\textendash Weiss law in the $T$-range 100\textendash 300
K. The inset shows the low-$T$ data down to 500 mK. (b) The data
of $C_{P}/T$ in zero-field. The down-arrow indicates $T_{N}$, while
the up-arrow points to the broad maximum ($T$$^{max}$).}
\end{figure}

Magnetization ($M$) as a function of temperature ($T$) is measured
on the single crystal in $H$ parallel to ($0ll$) plane. The magnetic
susceptibility $\chi=(M/H)$ in the $T$-range 500 mK to 300 K is
shown in Fig. \ref{2ChiCp}(a). At high-$T$, the data follow the
Curie-Weiss law with an effective magnetic moment $(\mu{}_{eff})\approx3.34\mu_{B}$
and a Curie-Weiss temperature $\theta{}_{CW}\approx-25$ K. The obtained
$\mu_{eff}$ value is larger than the expected value for $S=1$ (2.83
$\mu_{B}$), but is consistent with many Ni-based magnets \cite{NiBasedssystem,ArvindPRB(2007)}.
The obtained $\theta{}_{CW}$ of -25 K, indicates the presence of
antiferromagnetic couplings between the Ni$^{2+}$ ions. At low$-T$,
$\chi(T)$ shows a broad maximum around 16 K, indicative of short-range
spin correlations possibly originating from the low-dimensionality
of the system. Below the broad maximum, the susceptibility falls steeply
down to 1.4 K and then has an upturn. Unlike the spin-gap behavior
expected for isolated tetramer systems {[}20{]}, the data deviate
from the upturn at about 1.13 K, suggesting an antiferromagnetic transition
(see inset of Fig. \ref{2ChiCp}(a)). We have also measured the magnetization
in $H$ perpendicular to ($0ll$) plane, but no significant anisotropy
was seen. Specific heat $C_{P}(T)$ data measured on a single crystal
in zero-field are shown in Fig. \ref{2ChiCp}(b). The data of $C_{P}/T$
versus $T$ show features similar to those observed in $\chi(T)$:
a broad maximum at $T^{max}$ $\approx$ 5 K and a sharp transition
at $T_{N}\thickapprox1.13$ K. The observed $T^{max}$ of $C_{P}(T)$
is smaller than that of $\chi(T)$, as observed in other low-dimensional
spin systems \cite{refBiCu2PO6,BKoteswararaoSrCuTe2O6}. The appearance
of a sharp peak at $T_{N}$ infers the presence of LRO possibly due
to non-negligible inter-tetramer interactions. 

\begin{figure}
\begin{centering}
\includegraphics[clip,scale=0.32]{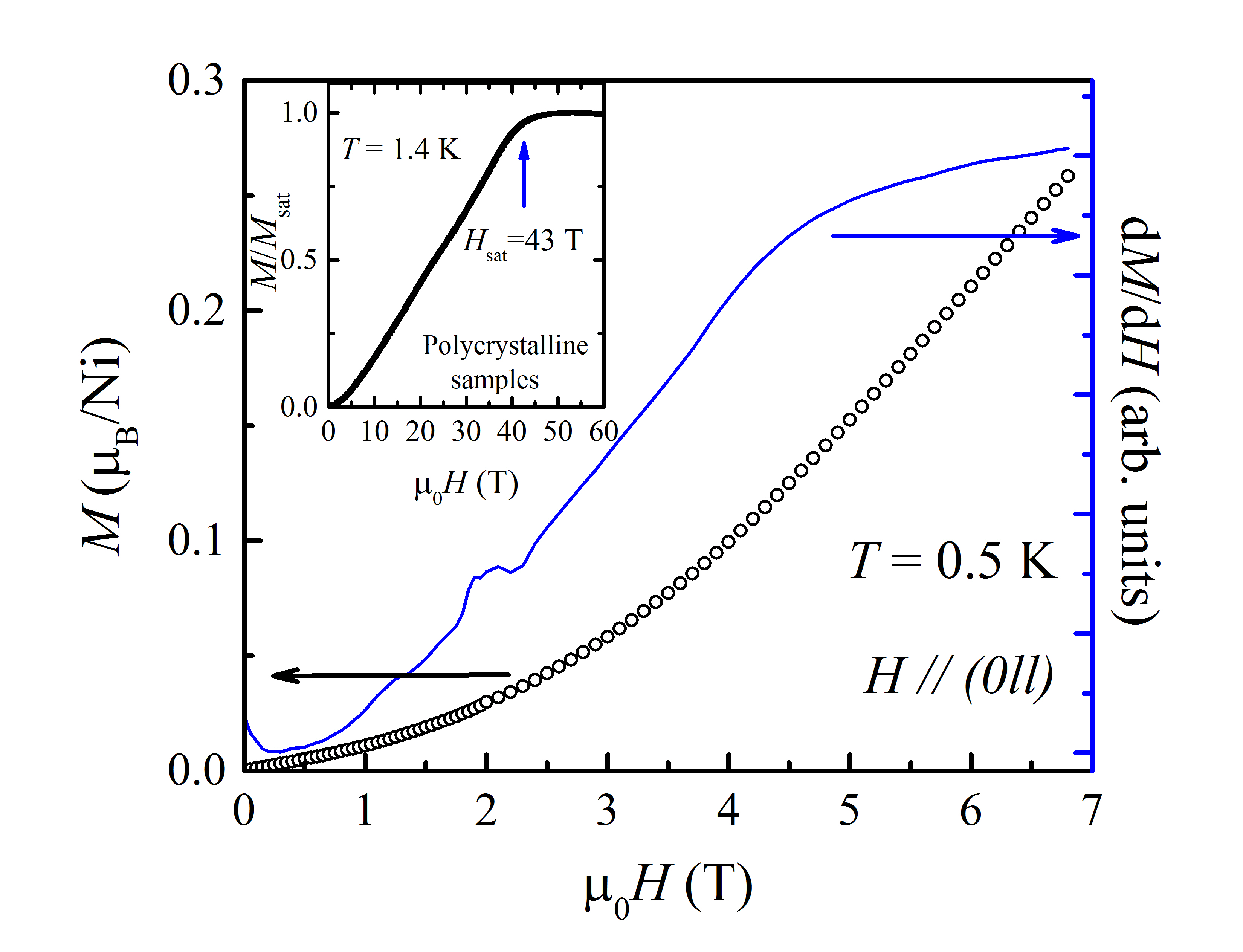}
\par\end{centering}

\caption{(Color online) \label{3MH data} (a) $M(H)$ up to 7 T at 0.5 K (left
$y$-axis). d$M$/d$H$ versus $H$ is plotted on the right $y$-axis.
Inset shows $M$($H$) data up to 60 T on a polycrystalline sample.}
\end{figure}

To explore further the nature of magnetic phenomena of this quantum
magnet, magnetization isotherm $M$($H$) was measured up to 7 T at
$T$ = 0.5 K (<$T$$_{N}$) as shown in Fig. \ref{3MH data}. $M$($H$)
data do not exhibit any hysteresis, ruling out the presence of ferromagnetic
moment. In addition, the data show a non-linear behavior, unlike in
a typical antiferromagnetic system. Similar non-linear behavior is
also seen in low-fields ($H$ < 5 T) in the $M$($H$) measured upto
60 T on the polycrystalline sample at 1.4 K, i.e., in the paramagnetic
region ($T$>$T$$_{N}$), as shown in the inset of Fig. \ref{3MH data}.
The $M$($H$) data suggest the appearance of field-induced phenomenon
in this quantum magnet. The magnetization increases with $H$ and
finally a fully polarized state with a saturated magnetization ($M$$_{sat}$)
about 2 $\mu_{B}$ /Ni ($M/M_{sat}$ = 1) is observed beyond $H$$_{sat}$
= 43 T.

\begin{figure}
\begin{centering}
\includegraphics[clip,scale=0.4]{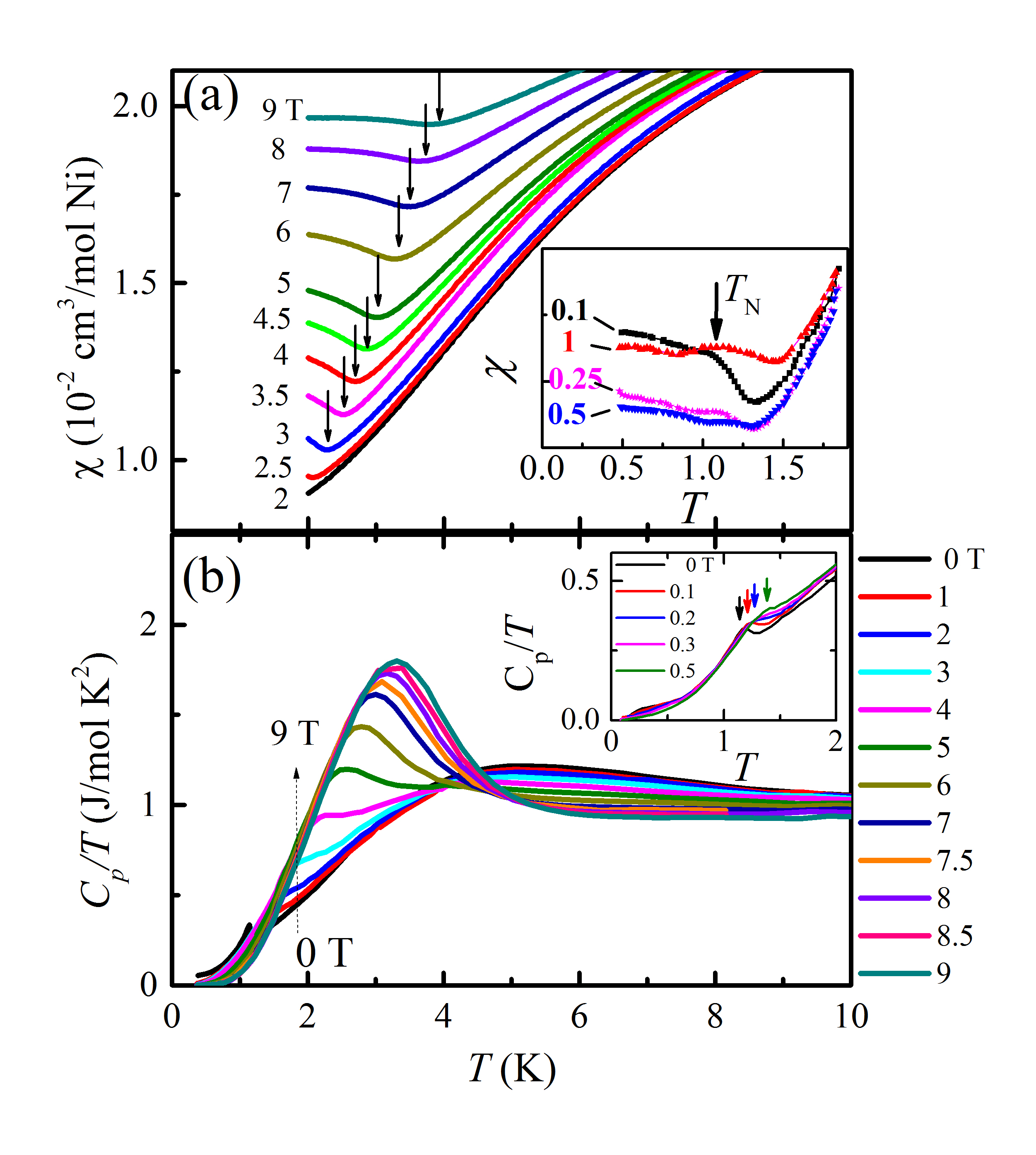}
\par\end{centering}

\caption{(Color online) \label{4ChiCpinHs} (a) $\chi(T)$ data at 2 T to 9
T. The minima in $\chi(T)$ depicts a phase transition which could
be reconciled as BEC of magnetic excitations. (b) $C_{P}/T$ versus
$T$ for the fields from 0 T to 9 T. Inset shows the data at low-fields.
The field-induced anomalies are represented by down arrows. }
\end{figure}

\begin{figure}[h]
\begin{centering}
\includegraphics[scale=0.54]{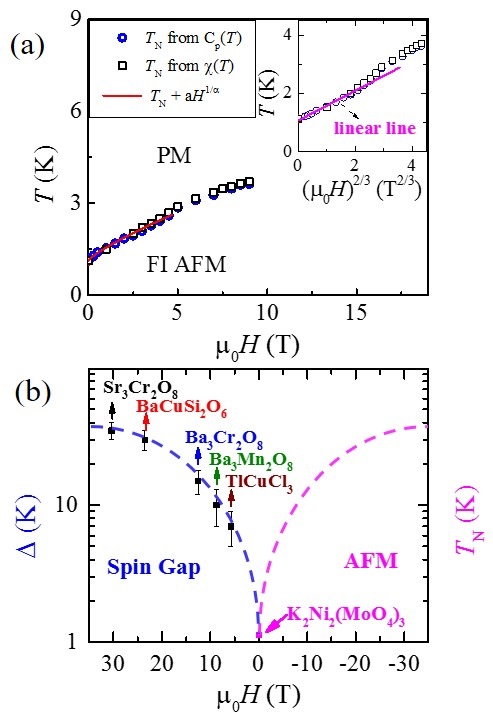}
\par\end{centering}

\caption{(Color online) \label{5PD} (a) $T-H$ phase diagram of K$_{2}$Ni$_{2}$(MoO$_{4}$)$_{3}$.
The red solid line is the fit mentioned in the text. Inset: the plot
of $T$$_{N}$ versus $H$$^{2/3}$. The pink line is an indication
of linear behavior. (b) Schematic quantum phase diagram with $H$
as tuning parameter. A QCP separates the spin-gap and the AFM-region.
Some selected spin-gap compounds are on the left side of the phase
diagram, while K$_{2}$Ni$_{2}$(MoO$_{4}$)$_{3}$ is positioned
on the AFM-side.}
\end{figure}

In order to understand the nature of field-induced phenomena in this
material, we measured $\chi(T)$ on a single crystal in the $T$-range
2-300 K and $C_{P}(T)$ down to 70 mK under fields up to 9 T. As shown
in the inset of Fig.\textbf{ \ref{4ChiCpinHs}}(a), a small $H$ of
0.25 T suppresses the $\chi(T)$ anomaly at $T_{N}$. On further increasing
$H$, surprisingly, the $\chi(T)$ data move to higher-$T$ and exhibit
dip-like anomalies, which has been observed, so far, in several spin-gap
materials exhibiting the field-induced BEC of triplons (when $H$
> $H$$_{C}$) \cite{T. Giamarchi NP2008,Vivien Zapf RMP(2014)}.
The dip-like anomalies or minimum in $\chi(T)$ were also evidenced
by theoretical simulations to support the BEC state of triplons \cite{T. NikuniPRL2000}.
Similarly, a cusp-like broad anomaly is observed in $C_{P}/T$ data
under a small $H$ of 0.1 T (see inset of Fig.\textbf{ \ref{4ChiCpinHs}}(b)).
The anomaly also moves to higher temperature with increasing $H$.
The observed transition and field-induced anomalies ($T$$_{FI}$)
from $\chi(T)$ and $C_{P}(T)$ data at different fields are plotted
in Fig. \ref{5PD} (a), which separates the field-induced antiferromagnetic
(FI-AFM) and paramagnetic (PM) regions. In order to evaluate the value
of the critical exponent, the phase boundary is fitted with the equation
$T_{FI}=T_{N}+aH{}^{1/\alpha}$, where $a$ is a proportionality factor
and $\alpha$ is an exponent. As suggested in the Ref. \cite{ONohadaniPRB2004}
that it is to be fitted below $T<$ 0.4 $T$$_{FI}$$^{max}$ to get
the precise exponent value. Here, $T_{FI}$$^{max}$ is the maximum
temperature at which a field-induced transition can take place (the
plateau in the $T-H$ phase diagram). The obtained value of $\alpha$
is found to be 1.4 (1); rather close to the theoretical value of the
exponent 3/2 predicted for 3D BEC of universality class \cite{ONohadaniPRB2004,TGiamarchiPRB1999}.
Moreover, the $T$$_{FI}$ values almost vary linearly with $H$$^{2/3}$
(see inset of \ref{5PD}(a)). The observed dip-like anomalies and
the obtained $\alpha$ value suggest that the compound K$_{2}$Ni$_{2}$(MoO$_{4}$)$_{3}$
exhibits field-induced BEC behavior. It is interesting to notice that
a very small field (0.1 T) is strong enough to induce this behavior
in a system with zero-field LRO at 1.13 K.

To further understand the field-induced behavior, we have compared
the data of a few spin-gap materials from literature which exhibit
a BEC of triplons (see Table 1). So far, BEC behavior has been realized
mostly in quantum mechanical spin-gap systems of the topological nature
without any symmetry-breaking. Magnetic field acts as chemical potential
and drives the density of triplons. As per the value of spin-gap and
critical-fields, we have positioned them in the phase diagram in Fig.\textbf{
}\ref{5PD}(b). The magnetic field acts as the tuning parameter and
it drives the spin-gap ground state to the AFM state via the quantum
critical point (QCP) at $T\rightarrow0$ K. It can be seen that all
the existing spin-gap materials are away from the QCP as a large value
of $H$ is required to suppress the spin-gap and finally to realize
the field-induced phenomena. On the other hand, if any system is in
the proximity of QCP then either the spin-gap and/or $T_{N}$ approaches
zero, as shown in Fig.\textbf{ }\ref{5PD}. In such a case, a small
critical $H$ would be sufficient to perturb its state. As we have
already observed that a small $H$ induces behavior akin to BEC of
triplon excitations, we conclude that K$_{2}$Ni$_{2}$(MoO$_{4}$)$_{3}$
is very close to the QCP on the AFM-side of the quantum phase diagram.
We would also like to discuss the possiblity that the ground state
of K$_{2}$Ni$_{2}$(MoO$_{4}$)$_{3}$ might have a mixture of singlets
and triplets (hence causes the LRO).Due to this reason, a small amount
of Zeeman energy is required for the triplon exciations. In general,
BEC corresponds to the spontaneous formation of a collective state
with macroscopic number of bosons governs by a single wave function.
In this antiferromagnet, the BEC of triplons state formed probably
due to the coherent precession of transverse magnetization, which
breaks U(1) symmetry, at zero and finite fields. Whatsoever the origin
of this unusual phenomena, the titled system experiences quantum mechanical
fluctuations but orders at finite-$T$. It appears to be close to
the quantum critical state $i.e.$, an extremely small gap or transition
at low-$T$. The determination of coherence lengths via Inelastic
Neutron Scattering measurements at zero-field and applied magnetic
fields would be useful to further understand the BEC mechanism in
this material.

\begin{table}
\caption{Some details of spin-gap and antiferromagnetic materials.}

\centering{}%
\begin{tabular}{|c|c|c|c|c|}
\hline 
Compound  & Type  & $\Delta$ or $T$$_{N}$ (K) & $H$$_{c}$(T) & Ref.\tabularnewline
\hline 
\hline 
Sr$_{3}$Cr$_{2}$O$_{8}$  & $S=1/2$ dimer  & 35  & 30.4 & \cite{A A AczelPRL2009}\tabularnewline
\hline 
BaCuSi$_{2}$O$_{6}$ & $S=1/2$ dimer & 30 & 23.5 & \cite{Sebastian2005,MJaimePRL2004}\tabularnewline
\hline 
Ba$_{3}$Cr$_{2}$O$_{8}$ & $S=1/2$ dimer & 15 & 12.5 & \cite{A A Aczel2009}\tabularnewline
\hline 
Ba$_{3}$Mn$_{2}$O$_{8}$ & $S=1$ dimer & 10 & 8.7 & \cite{E. C. SamulonPRL2009}\tabularnewline
\hline 
TlCuCl$_{3}$ & $S=1/2$ dimer & 7 & 5.7 & \cite{T. NikuniPRL2000}\tabularnewline
\hline 
K$_{2}$Ni$_{2}$(MoO$_{4}$)$_{3}$ & $S=1$ tetramer & $T$$_{N}$ = 1.13  & \ensuremath{\ge} 0.1 & this work\tabularnewline
\hline 
\end{tabular}
\end{table}

We looked at several other magnetic models and the corresponding critical
points. In the case of $S=1/2$ dimers, the magnitude of the spin-gap
($\Delta$) is same as that of the exchange coupling ($J$) \cite{dcjohnstom}.
But, the presence of significant inter-dimer coupling $J'/J=0.7$
(which is at the QCP), as in the Shastry-Sutherland model, can stabilise
an AFM state \cite{qcp of dimer}. In the case of the $S=1/2$ spin-ladder,
the QCP is predicted to be at a relative strength $J\lyxmathsym{\textquoteright}/J=0.3$
\cite{Sachdev2000}. Hence, we believe that a relative inter-tetramer
strength in K$_{2}$Ni$_{2}$(MoO$_{4}$)$_{3}$ might have placed
it at QCP.\textbf{ }Moreover, the presence of $nnn$ intra-tetramer
couplings, which causes the magnetic frustration, does not seem to
be negligible as the NiO$_{6}$ units are coupled to each other $via$
MoO$_{4}$ units with the path Ni-O-Mo-O-Ni. The bond-angles of Ni-O-Mo,
O-Mo-O, and Mo-O-Ni are about 116, 113, and 140$^{\circ}$, respectively,
which usually favor AFM couplings (see Fig. \ref{1XRD=000026CRYST}
(d) \& (e)). Magnetic frustration which usually enhance quantum fluctuations,
perhaps also plays a crucial role in placing this antiferromagnet
near a QCP. Further theoretical models would help to estimate the
relative strength exchange couplings to understand the origin of quantum
critical behavior.

In summary, we have successfully grown single crystals of the $S=1$
tetramer system K$_{2}$Ni$_{2}$(MoO$_{4}$)$_{3}$ and investigated
magnetization and specific heat studies. $\chi(T)$ and zero-field
$C_{P}(T)$ reveal that K$_{2}$Ni$_{2}$(MoO$_{4}$)$_{3}$ exhibits
LRO at $1.1$3 K due to the possible involvement of non-negligible
3D couplings, in contrast to the spin-gap behavior expected for an
isolated tetramer system. However, a small $H$ of about 0.1 T induces
a change in the magnetic behavior. The field-induced transition temperature
increases with increasing $H$ and follows $H$$^{1/\alpha}$ behavior
with $\alpha=1.4(1)$, which suggests that the observed field-induced
phenomena might be related to the BEC of triplons as observed in other
spin-gap materials. Despite having LRO in zero-field, the field-induced
behavior even in low-fields might point towards condensation of triplon
excitations with the possibility that K$_{2}$Ni$_{2}$(MoO$_{4}$)$_{3}$
is located in the vicinity of a quantum critical point in the phase
diagram. The ground state might have a mixture of singlets and triplets,
due to which, a small $H$ could induce BEC excitations in this material.
We believe that our results will draw attention to explore more insights
into the quantum criticality of the titled material.

\textbf{\textit{Acknowledgements:}} B.K. thanks DST INSPIRE faculty
award-2014 scheme. F.C.C. acknowledges Ministry of Science and Technology
in Taiwan under project number MOST-102-2119-M-002-004. We thank Prof.
Thamizhavel for providing the facilities to measure specific heat
at TIFR, Mumbai. We acknowledge the support of the HLD at HZDR, member
of the European Magnetic Field Laboratory (EMFL). A.V.M. thanks DST
for financial support.

\end{document}